\documentclass{ws-procs975x65}

\begin{document}

\title{Effective Gravity and Homogeneous Solutions} 

\author{Daniel M\"uller$^*$}

\address{Instituto de F\'\i sica, Universidade de Bras\'\i lia,\\
Cxp 04455 Bras\'\i lia, DF, 70919-970, Brazil \\
$^*$E-mail: muller@fis.unb.br}

\begin{abstract}
Near the singularity, gravity should be modified to an effective theory, in the same sense as with the Euler-Heisenberg electrodynamics. This effective gravity surmounts to higher derivative theory, and as is well known, a much more reacher theory concerning the solution space. On the other hand, as a highly non linear theory, the understanding of this solution space must go beyond the linearized approach. In this talk we will present some results previously published by collaborators and myself, concerning solutions for vacuum spatially homogeneous cases of Bianchi types $I$ and $VII_A$. These are the anisotropic generalizations of the cosmological spatially``flat", and ``open" models respectively. The solutions present isotropisation in a weak sense depending on the initial condition. Also, depending on the initial condition, singular solutions are obtained. 
\end{abstract}


\bodymatter

\section{Introduction}

In its primordial stage, quantum interactions dominate the evolution of
the Universe. In an effective approach, gravitation behaves classically
while the other fields are quantized,  Ref. \refcite{UdW}. It is well known
that due to vacuum polarization Ref. \refcite{S}, in the presence of a very
intense quantized fermion field, electromagnetism is modified to
its non linear version of Heisenberg-Euler \cite{H-E}. In this context,
the own dynamics of the electromagnetic field $F_{\mu\nu}F^{\mu\nu}$
is mandatory in order to obtain a finite theory. 

In the gravitational case, the counterterms necessary in order to obtain a finite theory result in the following Lagrangian, Ref. \refcite{dw}, 
\begin{equation}
\mathcal{L}=\sqrt{-g}\left[-\Lambda+R+
\alpha\left(R_{ab}R^{ab}-\frac{1}{3}R^{2}\right)+\beta R^{2}\right],
\label{acao}
\end{equation}
where $R_{ab}R^{ab}-\frac{1}{3}R^{2}$ is dynamically equivalent to the square of the Weyl tensor and $\alpha$ and $\beta$ are constants. The reason it is called the Schwinger-de-Witt formalism.

It is well known that linearization over a background reveals a spin $0$ field with mass $m_0=\sqrt{1/(-6\beta)}$ with one degree of freedom, a spin $2$ massless field with two degrees of freedom, and a spin $2$ ghost field with mass $m_2=\sqrt{1/\alpha}$,  see Ref. \refcite{Chiba}. The tachyon can be excluded by appropriate choices  of $\alpha>0$ and $\beta<0$. In the linearized limit, and absence of sources $T_{ab}=0$, the above mentioned $8$ degrees of freedom are entirely free. Classically the ghost field introduces non linear instability.

This higher order theory was previously studied by Starobinsky
\cite{Sby}.  However, solutions were found previously in Ref. \refcite{Bu} It is of interest for example in the context of the final stages of evaporation of black holes,  inflationary theories, Ref. \refcite{coule}, in the approach to the singularity Ref. \refcite{montani} and also in a more theoretical context Ref. \refcite{cotsakis}. We refer the reader to a paper by Schmidt \cite{hjs} for a review on higher order gravity theories in connection to cosmology.

In this talk, only vacuum solutions, $T_{ab}=0$, are considered.  

\section{Quadratic gravity and spatially homogeneous spaces}
Metric variations in eq. (\ref{acao}) results in the following equation of motion 
\begin{equation}
E_{ab}\equiv G_{ab}+\frac{1}{2}g_{ab}\Lambda-\left(\beta-\frac{1}{3}\alpha\right
)H_{\: ab}^{(1)}-\alpha H_{\: ab}^{(2)}=0,\label{eq.campo}
\end{equation}
where
\begin{eqnarray*}
&&G_{ab}=R_{ab}-\frac{1}{2}g_{ab}R,\\
&&H_{ab}^{(1)}=\frac{1}{2}g_{ab}R^{2}-2RR_{ab}-2g_{ab}\square R+2R_{;ab},\\
&&H_{ab}^{(2)}=\frac{1}{2}g_{ab}R^{cd}R_{cd}-\square R_{ab}-
\frac{1}{2}g_{ab}\square R+R_{;ab}-2R^{cd}R_{cbda}.
\end{eqnarray*}
Let us emphasize that every Einstein space satisfying $R_{ab}=
g_{ab}\Lambda/2$ is an exact solution of \eqref{eq.campo}. 

Considering the general line element $ds^2=-dt^2+h_{ij}\omega^i\otimes \omega^j,$ and the geodesic time like vector $u^a=(1,0,0,0)$, the torsion free connection is the following 
\begin{eqnarray*}
&&K_{ij}=\nabla_i u_{j}=\Gamma^0_{ij}=\frac{1}{2}\dot{h}_{ij}\\
&&K_i^j=h^{j k}K_{k i}=\Gamma^j_{i 0}=\Gamma^j_{0i},\\
&&\Gamma_{ijk}=\frac{1}{2}\left( C_{ijk}-C_{jik}-C_{kij}\right),
\end{eqnarray*}
where $[e_i,e_j]=-C^k_{ij}e_k$ are the structure constants, and the $e_i$ are the dual to the left invariant $1-$ form basis $\omega^k$. The Riemann tensor, and the field equations follow from the above connection. 

\section{Bianchi $I$ solutions}
In this case the structure constants $C^i_{jk}=0,$ which implies that the ${}^3R^i_{jkl}\equiv 0$.  The left invariant $1-$ form basis is $\omega^1=dx,$ $\omega^2=dy$, $\omega^3=dz$. 

The diagonal metric was addressed in Refs. \refcite{sandro,daniel-sandro,daniel-marcio-jcarlos}. While the non diagonal metric was addressed in Refs. \refcite{daniel-juliano,daniel}. Depending on the initial condition, the solutions oscillate toward a de Sitter type solution. In Ref. \refcite{daniel-juliano}, we have specifically checked that the frequencies, in a WKB sense are in agreement  with the masses of the linearized fields $\omega_{2}^{\mbox{{\tiny grav.}}}=0$, $\omega_0=\sqrt{1/(-6\beta)}$ and $\omega_{2}^{\mbox{{\tiny ghost}}}=\sqrt{1/\alpha}$.

We found a de Sitter attractor in Ref. \refcite{daniel-sandro} in the sense that many initial conditions converge to this solution. We also found many initial conditions which converge to a singularity. In the diagonal case we also explicitly checked that the oscillations toward the isotropic case have a tensorial and a scalar component Ref. \refcite{daniel-marcio-jcarlos}.  Of course these components are the linearized fields mentioned in the Introduction. A decomposition $3+1$ is given in Ref. \refcite{daniel}. The Bianchi $I$ case should deserve more attention since it occurs whenever the $3-$ curvature ${}^3R^i_{jkl}\rightarrow 0$.

\section{Bianchi $VII_A$ solutions}
In this case the non zero structure constants are $C^2_{12}=C^3_{13}=A,\; C^2_{31}=C^3_{12}=1$, and the left invariant $1-$ form basis is $\omega^1=dx,$ $\omega^2=e^{Ax}(\cos (x)dy-\sin (x)dz)$ and $\omega^3=e^{Ax}(\sin (x)dy+\cos (x)dz)$.

The non diagonal case was addressed in \refcite{juliano-daniel}. Also here depending on the initial conditions, a de Sitter type solution is approached in an oscillatory fashion. Again we have specifically checked that $\omega_{2}^{\mbox{{\tiny grav.}}}=0$, $\omega_0=\sqrt{1/(-6\beta)}$ and $\omega_{2}^{\mbox{{\tiny ghost}}}=\sqrt{1/\alpha}$, are the WKB frequencies. Depending on the initial condition, the solution approaches a singularity also. The intention is to explicitly check the influence of the $3-$ curvature on the dynamics. 
 \section*{Acknowledgments}
 D. M. thanks DPP-UnB for financial support.
 
\bibliographystyle{ws-procs975x65}

\end{document}